\documentclass{article}

\usepackage{microtype}
\usepackage{graphicx}
\usepackage{subcaption}
\usepackage{booktabs} 

\usepackage{hyperref}

\usepackage[preprint]{icml2026}

\usepackage{amsmath}
\usepackage{amssymb}
\usepackage{mathtools}
\usepackage{amsthm}

\usepackage[capitalize,noabbrev]{cleveref}

\theoremstyle{plain}

\theoremstyle{definition}

\theoremstyle{remark}
\usepackage{array}
\usepackage{xurl} 
\usepackage{etoolbox}
\apptocmd{\thebibliography}{\raggedright}{}{}

\usepackage[textsize=tiny]{todonotes}

\icmltitlerunning{Small models,  big threats: Characterizing safety challenges from low-compute AI models}

\begin{document}

\twocolumn[
  \icmltitle{Small models,  big threats: Characterizing safety challenges from low-compute AI models}



  \icmlsetsymbol{equal}{*}

  \begin{icmlauthorlist}
    \icmlauthor{Prateek Puri}{yyy}
  \end{icmlauthorlist}

  \icmlaffiliation{yyy}{Department of Engineer and Applied Sciences, RAND, Arlington, VA, USA}

  \icmlcorrespondingauthor{Prateek Puri}{ppuri@rand.edu}

  \icmlkeywords{Machine Learning,  AI Safety,  AI Governance,  Large Language models}

  \vskip 0.3in
]



\printAffiliationsAndNotice{}  

\begin{abstract}

Artificial intelligence (AI) systems are revolutionizing fields such as medicine, drug discovery, and materials science; however, many technologists and policymakers are also concerned about the technology's risks. To date, most concrete policies around AI governance have focused on managing AI risk by considering the amount of compute required to operate or build a given AI system. However, low-compute AI systems are becoming increasingly more performant - and more dangerous. Driven by agentic workflows, parameter quantization, and other model compression techniques, capabilities once only achievable on frontier-level systems have diffused into low-resource models deployable on consumer devices. In this report, we profile this trend by downloading historical benchmark performance data for over 5,000 large language models (LLMs) hosted on HuggingFace, noting the model size needed to achieve competitive LLM benchmarks has decreased by more than 10X over the past year. We then simulate the computational resources needed for an actor to launch a series of digital societal harm campaigns - such as disinformation botnets, sexual extortion schemes, voice-cloning fraud, and others - using low-compute open-source models and find nearly all studied campaigns can easily be executed on consumer-grade hardware. This position paper argues that protection measures for high-compute models leave serious security holes for their low-compute counterparts, meaning it is urgent both policymakers and technologists make greater efforts to understand and address this emerging class of threats. 

\end{abstract}

\section{Introduction}
\label{introduction}

Artificial intelligence (AI) technologies are enabling the widespread automation of information processing and reasoning tasks. Many anticipate these technologies will herald an era of unprecedented human productivity and economic output, while others are concerned they may be weaponized to cause large-scale societal harm. For example, researchers have investigated the extent to which advanced models, specifically large language models (LLMs), may facilitate synthetic biology attacks, compromise cybersecurity systems, and amplify the effects of disinformation campaigns \citep{helmus2022ai, hendrycks2023catastrophic}. 

Numerous policy frameworks have been proposed to mitigate AI risks, many of which focus on monitoring or regulating access to compute \citep{sastry2024computing}. For example, current US semiconductor export controls are designed, at least partially, to prevent the misuse of advanced models requiring high performance graphics processing units (GPUs), with US officials citing threats from "both AI training and inference at scale" \citep{BIS2024ExportControls}. Similarly, The European Union (EU) AI Act designates 10$^{25}$ training floating point operations (FLOPs) as a threshold for systemic risk categorization and regulation \citep{EuropeanParliament2023EUAIAct}.

However, training or deploying large models is not the only pathway to dangerous capabilities. Advancements in test-time compute, parameter quantization, agentic workflows, LLM tooling, and other techniques are rapidly diffusing capabilities from large AI systems into compact models that are easily deployable on consumer devices \citep{subramanian2025smalllanguagemodelsslms, lang2024comprehensivestudyquantizationtechniques, li2024reviewprominentparadigmsllmbased, shen2024smallllmsweaktool}. 

\textit{\textbf{Position}: The swift compression of advanced AI capabilities into smaller, accessible, and easily deployable models poses significant security risks that current governance frameworks are not fully equipped to handle. We urgently encourage more researchers and policymakers to focus on developing innovative governance strategies specifically tailored to low-compute AI threats, as these threats are becoming increasingly severe, frequent, and difficult to detect.}

In this report, we quantitatively profile the rate at which open-source LLMs have become both more performant and more compute-efficient over time. Secondly, we outline how this shift has impacted the amount of compute resources needed for a single actor to execute a variety of societal harm campaigns. Next we profile the computational workloads of several academic and commercial AI use cases, demonstrating a high degree of overlap between the compute required by both. We discuss how this overlap, combined with the relatively modest amount of compute required to launch the studied societal harm campaigns, complicates existing AI risk mitigation frameworks centered around high-compute models. Lastly, we briefly highlight the promises and shortcomings of a set of proposed strategies for mitigating low-compute AI risks.

As a clarification on terminology, we will refer to low-compute AI models in this report as those with <30B parameters, as these systems are increasingly deployable on low-cost hardware through parameter quantization and other inference optimization techniques. We will also use the term model compression to succinctly refer to the diffusion of AI capabilities into low-compute systems, although we acknowledge this phrase may have different meanings elsewhere in the literature. 

\section{Models are becoming more advanced at smaller sizes}
\label{profile}

Market pressures are guiding models to become both more capable and more lightweight over time, while hardware improvements are reducing barriers to model deployment. We discuss both trends below while observing that if these trends continue to evolve, bad-faith and good-faith actors alike will be able to deploy increasingly sophisticated models with commonly accessible levels of compute.

\subsection{Model miniaturization} 

We download performance data from over 5,000 open-source LLMs hosted on the HuggingFace LLM leaderboard. Each model on the leaderboard has been evaluated against the Eleuther AI Language Model Evaluation Harness \citep{eleuther2024lm}, a suite of benchmark tests designed to probe language model abilities on diverse tasks such as common sense reasoning, mathematical abilities, and others. In this report, we define a LLM's aggregate model performance, $\alpha$, as the mean score on the IFEval, BBH, MATH, GPQA, MUSR, and MMLU-PRO benchmark tests included within the Harness. \footnote{$\alpha$ is bounded between [0\%,100\%], with 100\% denoting a perfect score}.

\begin{figure*}[t]
  \centering
  \includegraphics[width=\textwidth]{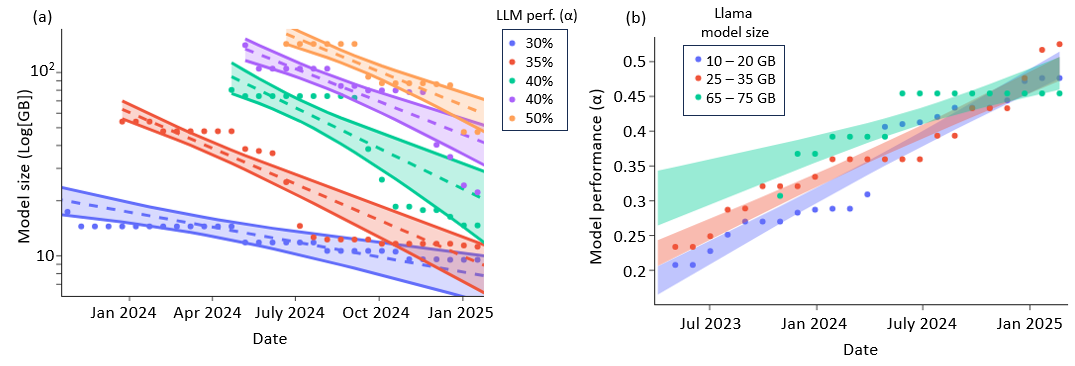}
  \caption{(a) Model size needed to obtain a given LLM benchmark score over time. Exponential curves are fit to the raw data and displayed along with the fit uncertainty bands. (b) Benchmark performance ($\alpha$) over time for three classes of Llama family models.}
  \label{fig1}
\end{figure*}

For each graded model on the leaderboard, we extract the model size (FP16 precision), model performance  ($ \alpha $), and the date on which the model was created (Appendix A). In Figure \ref{fig1}A, we plot the model size needed to obtain a given $\alpha$ over time. Each scatter point represents the 25$^{th}$ percentile model size value within the subset of models that surpass a given $\alpha$ at a given date. We present five different curves corresponding to $\alpha$ values of 30\%, 35\%, 40\%, 45\%, and 50\%. For reference, the highest $\alpha$ value listed on the HuggingFace leaderboard as of the writing of this report is 52\%.

We also fit a simple exponential decay curve to each set of data, which we present in the figure along with fit uncertainty bands for visual reference. As can be seen in Figure \ref{fig1}A, across all performance levels, the model size needed to obtain a given benchmark score has dropped significantly over time, with the model size needed to obtain $\alpha=0.35$ falling by $\sim$10X over the past year.\footnote{For robustness, we also recalculated these compression curves using each individual metric instead of the aggregate $\alpha$ value and observed similar compression trends over time.}

Similarly, Figure \ref{fig1}B displays how the $\alpha$ of a model of a fixed size has increased over time. Here, we filter our LLM leaderboard dataset to focus on Meta’s Llama family of models, given their active use within the developer community. As can be seen in the plot, Llama models of all size ranges \footnote{We cluster models into size ranges to account for the differing model sizes across the Llama 2, 3.1, 3.2, and 3.3 releases.} have steadily increased in performance over time and have even somewhat converged to a common $\alpha$ value of $\sim$45\%. Similar to Figure \ref{fig1}A, for visual reference, we fit linear models to the raw data and present them along with confidence intervals for each set of data. 

Put together, these plots suggest the following two trends: 
\begin{itemize}
\item The number of parameters needed to achieve a certain level of benchmark performance has decreased over time
\item The benchmark performance of a model of a given size has increased over time.
\end{itemize}

While these statements are hardly surprising, they have strong implications for both public safety and AI governance, which we will discuss further below. 

As a caveat, high performance on an LLM benchmark does not necessarily imply usefulness. In fact, many suspect that models have been engineered, or ‘overfit’, to provide deceptively high performance on such benchmarks while not providing high degrees of capability \citep{zhou2023dont}. We fully acknowledge the limitations of benchmark data, and later in this report, we will reference more in depth evaluations of the capabilities of low-compute models that have been conducted by other research groups. 

Similarly, while the above analysis focuses on open-source models, similar trends hold for closed-source models as well. For example, GPT-4 was offered at a price of \$120/million completion tokens \citep{wayback2024openai} in early 2023. However, GPT-4.1 – a model that performs comparably on many performance benchmarks while offering both longer context windows and multimodal abilities – now executes completions at roughly a 15X cheaper rate. Consequently, cost and resource barriers to deploying advanced AI capabilities are rapidly diminishing for malicious actors, regardless of whether they utilize open or closed-source models.

\subsection{Advances in hardware}

\begin{figure*}[t]
\centering
\includegraphics[width=1.0\linewidth]{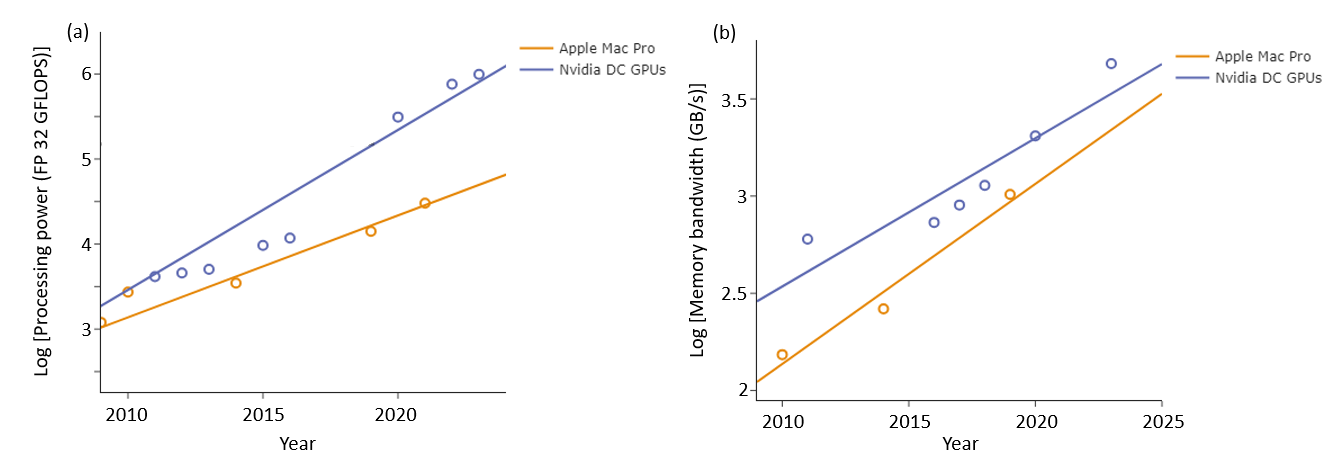}
\caption{(a) Evolution of processing power in NVIDIA and MacBook chips over time (b) Evolution of memory bandwidth across both sets of chips over time. Linear fits are presented alongside both curves, for visual reference.}
\label{fig2}
\end{figure*}

While high-performing models are becoming smaller, the processing power of accelerator chips is becoming larger, both across consumer and data center devices. In Figure \ref{fig2} we display the evolution of processing power (expressed in single-precision [FP32] FLOPS) and memory bandwidth (expressed in GB/s of memory transfer) over time for two sets of devices: NVIDIA data-center GPUs and consumer MacBooks GPUs.\footnote{We extracted performance metrics on both set of devices from official Apple and NVIDIA product pages.}

While many of the higher-performance data-center chips are currently restricted via US export controls, all MacBook chips explored in the chart are unrestricted and widely available for global use. These devices are sufficient to run inference on many advanced - and potentially dangerous - models, especially given the rapid pace of miniaturization in Figure \ref{fig1}a.

\section{The threats posed by low-compute AI systems}
\label{threat}

Building on the results of the previous section, we assess the risks of low-compute AI systems from three separate vantage points:

\begin{itemize}
\item Examining the rise in reported public AI security incidents
\item Highlighting studies evaluating the capabilities of low-compute AI systems
\item Simulating the compute required to launch AI-powered social harm campaigns
\end{itemize}

While each vantage point alone provides only a partial view of low-compute AI risk, together they collectively paint a more complete picture of the public security threats posed by these systems.

\subsection{The rapid rise of AI security incidents}

The FBI stated \$2.9B was lost through business-email-compromise scams in 2023 alone \citep{IC3_2024_GenerativeAI_Fraud, IC3_2023_AnnualReport}, citing GenAI as a key driver. Additionally, SlashNext reported a \$1,265\% increase in phishing incidents between Oct. 2022 and Sept. 2023 – a time period marked by the proliferation of generative AI technologies \citep{slashnext2023phishing}. Similarly, the FBI declared a “global sextortion crisis" fueled by generative models \citep{fbi2023sextortion}, while McAfee reported that 25\% of surveyed U.S. adults have either experienced or known someone who has experienced an AI voice scam \citep{McAfee2023ArtificialImpostor}.

While specific models used in these attacks are rarely disclosed, from the reporting that is available, we do see evidence of low-compute models. The original DeepNude app was a variant of a pix2pix model \citep{Cole2019DeepNude}, a lightweight system deployable on accessible hardware. Cloned versions of the original model still exist and are still contributing to the sex-tortion crisis mentioned above. In the future, attackers may leverage open-source, low compute models at even higher rates due to reduced surveillance and weaker content safeguards. For example, OpenAI recently flagged and shut down multiple high-profile accounts that were leveraging their models for cyberattacks, making low-compute offline systems even more attractive for bad-faith actors \citep{Nguyen2024ChatGPTCyberattacks}.

\subsection{Identification of dangerous capabilities within low-compute AI systems}

Researchers have tested the believability of audio, video, and text-based output from compressed models (<30B parameters) on human participants. For example, \citet{Hackenburg2025LLMpersuasion} demonstrated that open-source LLMs as small as 7B parameters were more politically persuasive than a human control group and equally persuasive to many larger LLMs. In \citet{Bray2023HumanDeepfakeDetection}, human participants were only able to identify deepfakes generated by a StyleGAN2 model (<1B-parameters) at a rate of $\sim$60\%, a value marginally above random chance. Similarly, in \citet{Warren2024AudioDeepfake}, in only $\sim$60\% of instances were human subjects able to detect synthetically generated audio samples within the WaveFake dataset \citep{Frank2021WaveFake}, a dataset consisting of audio samples generated via lightweight open-source models such as MelGAN ( <10M parameters). Lastly, \citet{Heiding2024SpearPhishing} and \citet{Schoenegger2025PersuasiveLLM} demonstrated that Claude-3.5 – which has been surpassed on the Chatbot Arena by multiple open-source LLMs <32B parameters in size - can produce spear-phishing emails that are as persuasive as those designed by human experts and is more effective at attitude, belief, and behavior shaping than a set of incentivized humans, respectively. 

\subsection{Simulating compute budgets of social harm campaigns}
\label{simulate}
Using the 2023 Executive Order on AI \citep{executive2023ai} as a guide, in this section we focus on a relevant set of disinformation, cybersecurity, voice cloning, and deepfakes threats.  We first performed a literature review to identify emblematic historical case studies of these types of attacks. Guided by the details of each case study, we decomposed each campaign into a set of constituent tasks executed over an associated timescale, and we subsequently estimated the amount of compute required for an AI model to replicate them. For example, a disinformation campaign can be decomposed into a sequence of generated social media posts, and a spear-phishing campaign may be broken down into a sequence of generated images and email chains.

Anchoring our analysis to historical case studies sets realistic scales for our simulated campaigns. For example, we could profile the compute load of a disinformation campaign consisting of 1,000 Tweets or 1,000,000 Tweets. These campaigns require vastly different compute requirements, and \textit{a-priori} it’s difficult to assess the societal harms posed by each. However, grounding our analysis in a historical disinformation case study helps both set the scale of a campaign and connect that scale to an event with understood social impact. For reference, we profile events like the Brexit disinformation campaign - an automated misinformation campaign thought to have influenced the outcome of a major geopolitical event \citep{Bruno2022} - and a business compromise scam that generated millions of dollars in company losses. For more specifics on each chosen historical case study, see Appendix \ref{appendix_c}.

\subsubsection{Simulation results}
\begin{figure*}
\centering
\includegraphics[width=0.6\linewidth]{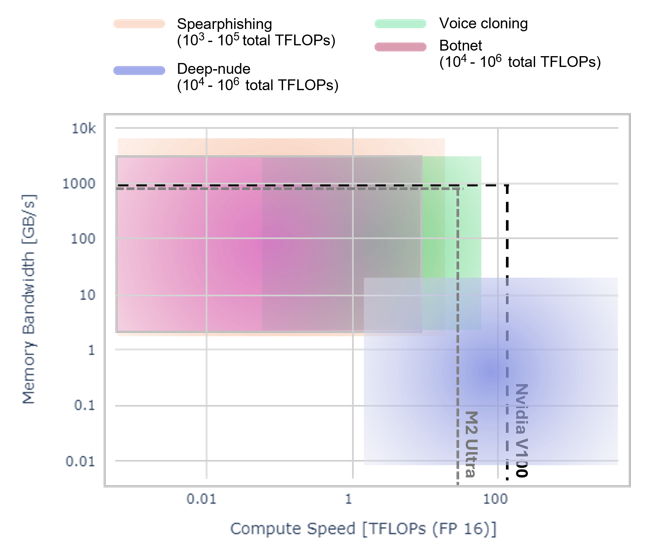}
\caption{The simulated compute profiles required to execute a set of disinformation, spearphishing, voice-cloning, and deepfake attacks with low-compute AI models. We break each attack into image, text, and audio generation steps and measure the memory speed and processing power an attacker would need to execute the attack using a single chip. The bounding boxes display the 5\% and 95\% for each GPU performance metric across our simulations. The dashed lines denote the performance metrics for NVIDIA V100 and Apple M2 Ultra chips, two currently non-export-controlled devices.}
\label{fig3}
\end{figure*}

\subsubsection{The periodic table of synthetic media attacks}
To estimate the compute required for each audio, text, and image generation task, we simulate each on a NVIDIA V100 GPU equipped with the nvprof GPU profiler\footnote{We profiled each generative model in half precision (FP16) format. Further, we validated our profiler by comparing the measured compute profiles of an LLM token generation task and matrix multiplication task to well-established theoretical estimates, observing largely consistent values between the two (Appendix \ref{appendix_b})}. Using this ‘periodic table’ of synthetic media generations, we build an aggregate compute profile of each campaign by considering the compute requirements of each constituent step. 

Of course, rather than measuring the compute profile for each task, one could also theoretically derive this profile. For example, both the memory and FLOP requirements of LLM token generation have well known analytic forms \citep{hoffmann2022empirical}. Similar analytic equations could, in theory, be extracted for image and audio models; however, the complexities of considering different batch sizes, image dimensions, etc. make measurement via the profiler more straightforward than theoretical derivation. Further, in later sections of this report, we profile the workloads of an entirely different set of non-nefarious AI workloads such as deep learning recommendation engine training where theoretical derivation is even less straightforward. 

\subsubsection{Estimating uncertainties}
Several variables impact the amount of compute needed within each campaign, such as the size of model required, the image resolution needed with a deepfake campaign, the number of seconds within a voice-clone scam voicemail, and the number of tokens generated within a botnet social media post. While historical case studies help constrain these parameters to some extent, they do not fix them entirely. To this end, we run Monte-Carlo simulations across these parameter spaces to generate compute profile uncertainties. We provide more details on the sampling framework leveraged within each campaign in Appendix \ref{appendix_d}.

In Figure \ref{fig3}, we plot the amount of memory bandwidth and compute speed required for an actor to execute each campaign on a single accelerator chip, and we also provide the total aggregate TFLOP of the subset of campaigns with fixed time-scales. Denoted by dashed lines, we display the bandwidth and processing power of the V100 and MacPro M2 Ultra accelerators – both of which are currently non-export controlled. 

A large fraction of the uncertainty boxes for all considered attacks are contained by the bounding boxes of the V100 and M2 Ultra accelerators, indicating such campaigns could in theory be executed with non-export-controlled devices. While certain regions of the uncertainty boxes lie beyond the performance bounds of a single chip, a majority of the studied campaigns are straightforward to distribute across multiple devices, meaning multiple chips could be combined to provide greater computing power. By our estimates, a computing cluster consisting of just ten V100 chips would offer enough computational power to surpass the estimated compute upper bounds of all considered threat campaigns - a system that could be purchased for mere thousands of dollars on eBay as of the writing of this report. Additionally, we intentionally adopted conservative simulation assumptions; in practice, attackers might achieve these outcomes with significantly fewer resources. 

Of course, these results have several limitations. We performed all profiling assuming generative models less than 30B parameters in size. Despite the studies referenced earlier, It is yet not totally proven that AI models of this size are performant enough to successfully execute the explored campaigns. However, if such capabilities do not yet exist at these model sizes, given the findings of Section \ref{profile}, they likely will soon. In general, it seems reasonable to assume the compute profile boxes in Figure \ref{fig3} will shift down and leftwards as models continue to become both more performant and compute efficient, reducing the compute resources needed to execute AI-powered attacks.

\section{Can't compute thresholds simply be adjusted to address threats from low-compute AI?}
\label{non_nefarious_profile}

The previous section demonstrated that current compute metering frameworks still permit access to compute levels sufficient to execute several societal harm campaigns. A natural follow-up question is: can these frameworks simply be revised to address this? The answer is complicated by the competing needs for these measures to both protect against AI risks while supporting non-nefarious business and academic development use cases. For example, export controls that successfully restrict bad actors from executing societal harm campaigns but also disrupt compute flows within AI-dependent industries would harm hardware manufacturers, strain international relationships, and hinder research collaborations. 

To address this point, similar to Section \ref{simulate}, we profile the compute required to execute a set of typical business/academic AI workloads and compare the results to those in the previous section. 

The selected workloads include the following: object recognition within autonomous vehicles, protein structure prediction within biomedical research, audio-to-text transcription within customer call centers, spam detection modeling, and recommendation engine training. We selected this set of workloads based on two criteria. First, we performed a literature review of most common AI commercial and academic use cases. Secondly, we filtered the identified set of workloads to diversify both across model type and application sector. For example, we include a recommendation engine example since - at least as of 2019 - these systems were estimated to be one of the highest volume workloads in global data centers \citep{mudigere2022softwarehardware}. Similarly, we diversified our workloads to include audio, video, and text generation models leveraged across both academic and commercial sectors.

\begin{figure}
\centering
\includegraphics[width=1.0\linewidth]{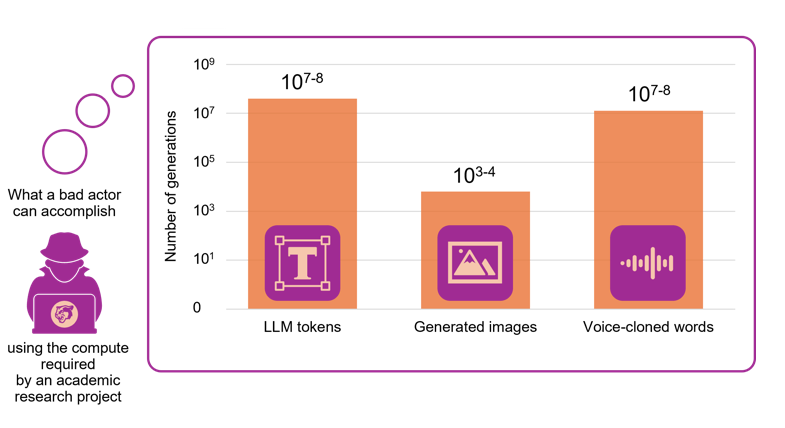}
\caption{The number of synthetic images, llm-generated tokens, and words of voice cloned audio an actor could generate with the compute required by a single typical academic experiment.}
\label{non_nefarious_fig}
\end{figure}

Following a procedure similar to that used above, we find the compute required by each workload exceeds that required within all profiled societal harm campaigns in Section \ref{simulate}, with more complete details provided in Appendix \ref{appendix_e}. This suggests that naively adjusting compute metering thresholds to block attacks from miniaturized AI systems would significantly disrupt many non-nefarious academic and business AI use cases. As an illustrative example, in Figure \ref{non_nefarious_fig}, we display the number of synthetic images, LLM tokens, and voiced-cloned words an actor could generate with the compute required by our biomedical research example. As can be seen in the plot, an actor could generate hundreds of millions of pieces of synthetic content with the compute required by a single academic experiment. 

\section{Alternative strategies for addressing low-compute AI risks}
\label{alternatives}

Collectively, the insights uncovered in the last two sections highlight the need for AI protection paradigms that are fundamentally different from existing approaches. In this section, we explore several proposed governance strategies and societal adaptations for addressing risks from low-compute AI while also acknowledging their implementation challenges.

\textbf{\textit{Capability-based risk evaluations:}} Rather than inferring AI risk from compute, alternative governance frameworks can focus on evaluating AI systems through their demonstrated capabilities. A capability-focused approach would evaluate systems based on benchmarks specifically designed to probe for potentially harmful abilities, such as persuasiveness, deception capabilities, or proficiency at tasks with dual-use potential, with flagged systems being prioritized for more extensive review and/or access restrictions \citep{shevlane2023modelgrading, Hooker2024ComputeThresholds}.

However, capability-based frameworks face significant implementation challenges. Creating robust, comprehensive capability evaluations requires substantial technical expertise, and benchmarks can quickly become outdated as the field rapidly advances \citep{amodei2023frontier}. Additionally, bad-faith developers may optimize their systems to perform poorly on regulatory benchmarks while maintaining harmful capabilities in deployment contexts. These challenges suggest the need for a technically adept regulatory body with sufficient resources to continuously update evaluation methodologies, an entity which will be nontrivial to establish and maintain. Nonetheless, capability-based frameworks can complement compute-based regulation by addressing the fundamental limitation that compute alone is an increasingly limited proxy for AI risk \citep{tamkin2023measuring}.

\textbf{\textit{Defensive AI:}} The offense-defense dynamic is well known in fields such as cybersecurity where automated spam detection and spam creation models have been co-evolving for decades. Extending this concept, one could imagine voice clone detection models thwarting attacks from voice cloning agents, automated cybersecurity agents patching vulnerabilities exposed by LLM-powered hackers, in addition to other offense-defense dynamics \citep{lohn2025impactaicyberoffensedefense}.

However, while defensive models have shown promise for defending against digital media and cybersecurity attacks, they may be ill-equipped to address other classes of threats with less favorable ‘offense-defense’ balances \citep{unver2023strategic, aspen2024cyber}. For example, even if public institutions have major compute advantages, it may be much easier to exploit offensive AI to design and deploy a bioweapon than to deploy defensive AI to produce and disseminate a vaccine amongst a population. 

Additionally, even if defense has theoretical advantages over offense in certain domains, firms may not have enough resources to implement solutions, especially given limited resources and talent. On one hand, regulation could at least provide firms clarity on what defensive protocols are necessary and relevant to implement; however such policies may also produce frictions that hinder AI innovation \citep{Yeung2025StepsAI}.

\textbf{\textit{Inference time-filtering and watermarking:}} If widespread access to advanced models is inevitable, integrating protective measures directly into models offers one approach to safety. Content filters can block harmful outputs, while digital watermarks embedded within AI-generated content \citep{roman2024proactive, deepmind2023synthid, kirchenbauer2023watermark} can help recipients identify synthetic material. For example, watermarks could alert individuals when they are speaking with a voice cloned agent, consuming content produced by a LLM, or being solicited by a generative AI agent. 

However, these approaches face substantial challenges. For example, generated content that is not blatantly toxic may still be socially dangerous. Content-wise, the script for a phone-based financial fraud scam and a benign telemarketing campaign may sound very similar, and toxicity filters alone may struggle to distinguish between the two. Secondly, researchers have demonstrated that both open-source and closed-source models can be coaxed into producing harmful outputs through straightforward prompting strategies, even when moderation safeguards are in place \citep{yu2023gptfuzzer}.

\textbf{\textit{Developing social AI resiliency:}} Embracing an preventative approach to managing AI risk, public officials could invest in making domestic institutions, infrastructure, and social groups more resilient \citep{bernardi2024societal} to AI attack. For example, social adaptations such as enhanced media literacy training, efficient AI incident reporting, and increased AI education can mitigate the impact of AI-powered attacks and make individuals less prone to AI deception. Similarly, AI models do not operate in a vacuum and often must be connected to other inputs/resources to execute tasks. For example, even if a LLM can generate the molecular structure of a bio-virus, actors need to access a laboratory environment to synthesize it.  As part of an AI resiliency strategy, officials can identify the raw materials and datasets needed for various AI attacks and built robust safeguards around them. 

However, once again, infrastructure and collaborations need to be built - and funded - before resiliency efforts can scale. Similarly, while safeguarding access to key datasets and materials may be effective in theory, identifying and rectifying critical points of vulnerability is a formidable challenge and will require coordination across many disparate sectors. 
\section{Alternative Views}

In this section, we present a set of alternative viewpoints that question or complicate the position of this paper while also providing counterpoints for each alternative viewpoint. 
 
\textbf{Compute-based policies will at least prevent the scaling of harmful campaigns}:
Limitations aside, compute-based policies can still limit the ability of an actor to scale attacks in cases where attack surface area scales meaningfully with compute. However, scale doesn't always translate clearly to impact. For example, the misinformation campaign simulation in Section \ref{threat}, while seemingly modest in size, matches the scale of a well-documented Brexit campaign ($\sim$50,000-100,000 posts) thought to have influenced a major geopolitical event. Similarly, an attacker who leverages AI to design a deadly pathogen might require only minimal compute resources but still cause catastrophic harm.

For certain threats (automated drone swarms, mass surveillance, etc.), scale certainly does directly increase severity. Compute metering policies will help mitigate the impact of these types of attacks while still rendering the public vulnerable to others, especially as the model miniaturization and hardware improvement trends in Section \ref{profile} continue to advance. 

\textbf{Public safety risks from low-cost compute AI models are already adequately addressed by existing AI governance approaches}: In the report we highlighted several examples where low-compute models either have been or can be exploited to cause social harm. All of the considered models would both fall under the EU AI Act's systemic risk compute threshold and are deployable on non-export-controlled hardware. Compute metering policies - while effective in many ways - still have vulnerabilities that need to be taken seriously. 

\textbf{Market forces will drive firms naturally to address low-compute AI threats, meaning policy solutions should focus on high-compute threats}: While market incentives theoretically encourage firms to mitigate AI security risks, recent surveys reveal organizations are struggling to address AI threats \citep{ArizeAI2024GenAISEC}, often due to a lack of resources, expertise, or incentives/requirements to implement adequate safeguards \citep{Netwrix2024HybridSecurity}. Our position is not that policy must solely solve low-compute AI challenges, but rather that existing risk mitigation approaches contain fundamental blind spots that can be rectified by strategic public-private collaborations.

\section{Conclusion}
\label{conclusion}

We explore how capabilities initially only present within larger-scale LLMs have diffused into low-resource, lightweight systems deployable on consumer devices. By analyzing historical performance data from over 5,000 models on HuggingFace, we demonstrate that the size of model needed to achieve competitive LLM benchmark scores has decreased by as much as 10X over the past year. We also simulate the compute needed for a bad-faith actor to launch a set of social harm campaigns, unveiling that many such attacks are easily executable on consumer hardware. 

While these trends have been noted by other researchers \citep{bommasani2023foundation, weidinger2022taxonomy, Hooker2024ComputeThresholds}, current AI governance frameworks have not adequately evolved to address the risks posed by increasingly powerful low-compute models. We present empirical evidence to underscore the urgency of rethinking approaches that rely primarily on compute thresholds as proxies for risk. As model miniaturization accelerates, policymakers must develop more nuanced frameworks that consider capabilities, intent, and potential for harm alongside compute requirements. This will require deeper collaboration between technical experts, policy makers, and industry to develop more comprehensiveness protection frameworks that consider a wider class of risks, rather than over-indexing on high-compute threats. The pace of AI advancement demands that these conversations move beyond mere discourse into concrete adaptations that meet today's rapidly evolving technological landscape.

\bibliography{example_paper}

@article{Hooker2024ComputeThresholds,
  author = {Sara Hooker},
  title = {On the Limitations of Compute Thresholds as a Governance Strategy},
  journal = {arXiv preprint arXiv:2407.05694},
  year = {2024},
  doi = {10.48550/arXiv.2407.05694},
  url = {https://arxiv.org/abs/2407.05694}
}

@techreport{IC3_2023_AnnualReport,
  title        = {2023 Internet Crime Report},
  institution  = {Federal Bureau of Investigation, Internet Crime Complaint Center (IC3)},
  year         = {2024},
  month        = {March},
  url          = {https://www.ic3.gov/AnnualReport/Reports/2023_IC3Report.pdf},
  note         = {Report released in March 2024, covering data from calendar year 2023}
}

@misc{lang2024comprehensivestudyquantizationtechniques,
      title={A Comprehensive Study on Quantization Techniques for Large Language Models}, 
      author={Jiedong Lang and Zhehao Guo and Shuyu Huang},
      year={2024},
      eprint={2411.02530},
      archivePrefix={arXiv},
      primaryClass={cs.LG},
      url={https://arxiv.org/abs/2411.02530}, 
}

@misc{li2024reviewprominentparadigmsllmbased,
      title={A Review of Prominent Paradigms for LLM-Based Agents: Tool Use (Including RAG), Planning, and Feedback Learning}, 
      author={Xinzhe Li},
      year={2024},
      eprint={2406.05804},
      archivePrefix={arXiv},
      primaryClass={cs.AI},
      url={https://arxiv.org/abs/2406.05804}, 
}

@misc{shen2024smallllmsweaktool,
      title={Small LLMs Are Weak Tool Learners: A Multi-LLM Agent}, 
      author={Weizhou Shen and Chenliang Li and Hongzhan Chen and Ming Yan and Xiaojun Quan and Hehong Chen and Ji Zhang and Fei Huang},
      year={2024},
      eprint={2401.07324},
      archivePrefix={arXiv},
      primaryClass={cs.AI},
      url={https://arxiv.org/abs/2401.07324}, 
}

@article{amodei2023frontier,
  title={Frontier AI Regulation: Managing Emerging Risks to Public Safety},
  author={Amodei, Dario and Anthropic Team and others},
  journal={Anthropic Technical Report},
  year={2023},
  url={https://www.anthropic.com/news/frontier-ai-regulation}
}

@misc{subramanian2025smalllanguagemodelsslms,
      title={Small Language Models (SLMs) Can Still Pack a Punch: A survey}, 
      author={Shreyas Subramanian and Vikram Elango and Mecit Gungor},
      year={2025},
      eprint={2501.05465},
      archivePrefix={arXiv},
      primaryClass={cs.CL},
      url={https://arxiv.org/abs/2501.05465}, 
}

@article{tamkin2023measuring,
  title={Measuring Progress in Large Language Models},
  author={Tamkin, Alex and Brundage, Miles and Clark, Jack and Ganguli, Deep},
  journal={arXiv preprint arXiv:2303.08774},
  year={2023}
}

@article{bommasani2023foundation,
  title={Foundation models and the future of AI: A new paradigm and emerging risks},
  author={Bommasani, Rishi and Zhang, Kathleen and Kobren, Ari and Bailis, Peter and Duchi, John and Hernandez, Danny and Horvitz, Eric and Jaakkola, Tommi and Kaplan, Jared and Koh, Pang Wei and others},
  journal={arXiv preprint arXiv:2307.13346},
  year={2023}
}

@article{weidinger2022taxonomy,
  title={A taxonomy of AI risks},
  author={Weidinger, Laura and Gabriel, Iason and Uesato, Jonathan and Mellor, Joseph and Goldie, William and Irving, Geoffrey and Sodhani, Shaona and Hendricks, Lisa Anne and Leike, Jan and Kasirzadeh, Atoosa and others},
  journal={arXiv preprint arXiv:2206.03421},
  year={2022}
}

@article{shevlane2023modelgrading,
  title={Model evaluation for extreme risks},
  author={Shevlane, Toby and Fleming, Michael and Hogarth, Rafe and Hadfield, Gillian},
  journal={arXiv preprint arXiv:2305.15324},
  year={2023}
}

@techreport{IC3_2024_GenerativeAI_Fraud,
  title        = {Criminals Use Generative Artificial Intelligence to Facilitate Financial Fraud},
  institution  = {Internet Crime Complaint Center (IC3)},
  type         = {Public Service Announcement},
  number       = {I-120324-PSA},
  year         = {2024},
  month        = {December},
  day          = {3},
  url          = {https://www.ic3.gov/PSA/2024/PSA241203}
}

@article{Heiding2024SpearPhishing,
  author = {Fred Heiding and Simon Lermen and Andrew Kao and Bruce Schneier and Arun Vishwanath},
  title = {Evaluating Large Language Models' Capability to Launch Fully Automated Spear Phishing Campaigns: Validated on Human Subjects},
  journal = {arXiv preprint arXiv:2412.00586},
  year = {2024},
  doi = {10.48550/arXiv.2412.00586},
  url = {https://arxiv.org/abs/2412.00586}
}

@article{Frank2021WaveFake,
  author = {Joel Frank and Lea Sch{\"o}nherr},
  title = {WaveFake: A Data Set to Facilitate Audio Deepfake Detection},
  journal = {arXiv preprint arXiv:2111.02813},
  year = {2021},
  note = {NeurIPS 2021 Datasets and Benchmarks Track},
  doi = {10.48550/arXiv.2111.02813},
  url = {https://arxiv.org/abs/2111.02813}
}

@article{Bruno2022,
  author = {Bruno, Matteo and Lambiotte, Renaud and Saracco, Fabio},
  title = {Brexit and bots: characterizing the behaviour of automated accounts on Twitter during the UK election},
  journal = {EPJ Data Science},
  volume = {11},
  number = {1},
  pages = {17},
  year = {2022},
  doi = {10.1140/epjds/s13688-022-00330-0},
  url = {https://doi.org/10.1140/epjds/s13688-022-00330-0},
  issn = {2193-1127}
}

@inproceedings{Warren2024AudioDeepfake,
  author = {Kevin Warren and Tyler Tucker and Anna Crowder and Daniel Olszewski and Allison Lu and Caroline Fedele and Magdalena Pasternak and Seth Layton and Kevin Butler and Carrie Gates and Patrick Traynor},
  title = {{``Better Be Computer or I'm Dumb'': A Large-Scale Evaluation of Humans as Audio Deepfake Detectors}},
  booktitle = {Proceedings of the 2024 ACM SIGSAC Conference on Computer and Communications Security (CCS '24)},
  year = {2024},
  address = {Salt Lake City, UT, USA},
  publisher = {Association for Computing Machinery},
  doi = {10.1145/3658644.3670325},
  url = {https://cise.ufl.edu/~butler/pubs/ccs24-warren-deepfake.pdf}
}

@techreport{Yeung2025StepsAI,
  author      = {Douglas Yeung and Tina Huang and Benjamin Boudreaux and Prateek Puri and Jonathan W. Welburn and Anita Chandra and Miriam Vogel},
  title       = {Steps Toward AI Governance: Insights and Recommendations from the 2024 EqualAI Summit},
  institution = {RAND Corporation},
  year        = {2025},
  type        = {Conference Proceedings},
  number      = {CFA3799-1},
  doi         = {10.7249/CFA3799-1},
  url         = {https://www.rand.org/pubs/conf_proceedings/CFA3799-1.html}
}

@article{Bray2023HumanDeepfakeDetection,
  author = {Sergi D. Bray and Shane D. Johnson and Bennett Kleinberg and Toby Davies and Lewis D. Griffin},
  title = {Testing human ability to detect 'deepfake' images of human faces},
  journal = {Journal of Cybersecurity},
  volume = {9},
  number = {1},
  pages = {tyad011},
  year = {2023},
  publisher = {Oxford University Press},
  doi = {10.1093/cybsec/tyad011},
  url = {https://doi.org/10.1093/cybsec/tyad011}
}

@article{Hackenburg2025LLMpersuasion,
  author = {Kobi Hackenburg and Ben M. Tappin and Paul R{\"o}ttger and Scott A. Hale and Jonathan Bright and Helen Margetts},
  title = {Scaling language model size yields diminishing returns for single-message political persuasion},
  journal = {Proceedings of the National Academy of Sciences},
  volume = {122},
  number = {10},
  pages = {e2413443122},
  year = {2025},
  publisher = {National Academy of Sciences},
  doi = {10.1073/pnas.2413443122},
  url = {https://doi.org/10.1073/pnas.2413443122}
}

@article{Nguyen2024ChatGPTCyberattacks,
  author  = {Nguyen, Britney},
  title   = {OpenAI found 'state-affiliated malicious actors' using ChatGPT for cyberattacks},
  journal = {Quartz},
  year    = {2024},
  month   = {February},
  day     = {14},
  url     = {https://qz.com/openai-microsoft-chatgpt-ai-cyberattacks-1851255460}
}

@article{Schoenegger2025PersuasiveLLM,
  author = {Philipp Schoenegger and Francesco Salvi and Jiacheng Liu and Xiaoli Nan and Ramit Debnath and Barbara Fasolo and Evelina Leivada and Gabriel Recchia and Fritz G{\"u}nther and Ali Zarifhonarvar and 
            Joe Kwon and Zahoor Ul Islam and Marco Dehnert and Daryl Y. H. Lee and Madeline G. Reinecke and David G. Kamper and Mert Koba{\c s} and Adam Sandford and Jonas Kgomo and Luke Hewitt and 
            Shreya Kapoor and Kerem Oktar and Eyup Engin Kucuk and Bo Feng and Cameron R. Jones and Izzy Gainsburg and Sebastian Olschewski and Nora Heinzelmann and Francisco Cruz and Ben M. Tappin and 
            Tao Ma and Peter S. Park and Rayan Onyonka and Arthur Hjorth and Peter Slattery and Qingcheng Zeng and Lennart Finke and Igor Grossmann and Alessandro Salatiello and Ezra Karger},
  title = {{Large Language Models Are More Persuasive Than Incentivized Human Persuaders}},
  journal = {arXiv preprint arXiv:2505.09662},
  year = {2025},
  doi = {10.48550/arXiv.2505.09662},
  url = {https://arxiv.org/abs/2505.09662}
}

@article{guo2025deepseek,
  title={{DeepSeek-R1}: Incentivizing Reasoning Capability in {LLMs} via Reinforcement Learning},
  author={Guo, Dongsheng and Yang, Dongxu and Zhang, Hui and Song, Jiaming and Zhang, Rui and Xu, Rui and Zhu, Qi and Ma, Shaokun and Wang, Peng and Bi, Xin and Zhang, Xiaodong},
  journal={arXiv},
  year={2025},
  month={jan},
  archivePrefix={arXiv},
  eprint={2501.12948},
  primaryClass={cs.AI},
  doi={10.48550/arXiv.2501.12948}
}

@inproceedings{kirchenbauer2023watermark,
  title={A Watermark for Large Language Models},
  author={Kirchenbauer, John and Geiping, Jonas and Wen, Yuxin and Katz, Jonathan and Miers, Ian and Goldstein, Tom},
  booktitle={Proceedings of the 40th International Conference on Machine Learning},
  series={Proceedings of Machine Learning Research},
  volume={202},
  pages={17061--17084},
  year={2023},
  month={jul},
  publisher={PMLR},
  url={https://proceedings.mlr.press/v202/kirchenbauer23a.html}
}

@misc{deepmind2023synthid,
  title={{SynthID}: Identifying {AI}-Generated Content with {SynthID}},
  author={{Google DeepMind}},
  year={2023},
  month={aug},
  howpublished={\url{https://deepmind.google/discover/blog/identifying-ai-generated-images-with-synthid/}},
  note={Accessed: 2025-05-16}
}

@techreport{helmus2022ai,
  title={Artificial Intelligence, Deepfakes, and Disinformation: A Primer},
  author={Helmus, Todd C.},
  institution={RAND Corporation},
  year={2022},
  month={jul},
  url={https://www.rand.org/pubs/perspectives/PEA1149-1.html}
}

@article{hendrycks2023catastrophic,
  title={An Overview of Catastrophic {AI} Risks},
  author={Hendrycks, Dan and Mazeika, Mantas and Woodside, Thomas},
  journal={arXiv},
  year={2023},
  month={jun},
  archivePrefix={arXiv},
  eprint={2306.12001},
  primaryClass={cs.AI},
  doi={10.48550/arXiv.2306.12001}
}

@inproceedings{hoffmann2022empirical,
  title={An Empirical Analysis of Compute-Optimal Large Language Model Training},
  author={Hoffmann, Jordan and Borgeaud, Sebastian and Mensch, Arthur and Buchatskaya, Elena and Cai, Trevor and Rutherford, Eliza and de Las Casas, Diego and Hendricks, Lisa Anne and Welbl, Johannes and Clark, Aidan and others},
  booktitle={Advances in Neural Information Processing Systems},
  volume={35},
  pages={30016--30030},
  year={2022}
}

@inproceedings{mudigere2022softwarehardware,
  title={Software–Hardware Co-Design for Fast and Scalable Training of Deep Learning Recommendation Models},
  author={Mudigere, Dheevatsa and Hao, Yuchen and Huang, Jianyu and Jia, Zhihao and Tulloch, Andrew and Sridharan, Srinivas and Liu, Xing and others},
  booktitle={Proceedings of the 49th Annual International Symposium on Computer Architecture},
  year={2022},
  doi={10.1145/3470496.3527409}
}

@inproceedings{dao2022flashattention,
  title={FlashAttention: Fast and Memory-Efficient Exact Attention with {I/O}-Awareness},
  author={Dao, Tri and Fu, Dan and Ermon, Stefano and Rudra, Atri and R{\'e}, Christopher},
  booktitle={Advances in Neural Information Processing Systems},
  volume={35},
  year={2022}
}

@article{yu2023gptfuzzer,
  title={{GPTFuzzer}: Red Teaming Large Language Models with Auto-Generated Jailbreak Prompts},
  author={Yu, Jiahao and Lin, Xingwei and Xing, Xinyu},
  journal={arXiv},
  year={2023},
  archivePrefix={arXiv},
  primaryClass={cs.CR}
}

@misc{eleuther2024lm,
  title={{LM} Evaluation Harness},
  author={{Eleuther {AI}}},
  year={2024},
  month={may},
  howpublished={\url{https://github.com/EleutherAI/lm-evaluation-harness}},
  note={Accessed: 2025-05-16}
}

@misc{executive2023ai,
  title={Executive Order on Safe, Secure, and Trustworthy Development and Use of Artificial Intelligence},
  author={Executive Order 14110},
  organization={Executive Office of the President},
  year={2023},
  month={oct}
}

@article{Cole2019DeepNude,
  author  = {Samantha Cole},
  title   = {This Horrifying App Undresses a Photo of Any Woman With a Single Click},
  journal = {VICE},
  year    = {2019},
  month   = {June},
  day     = {26},
  url     = {https://www.vice.com/en/article/deepnude-app-creates-fake-nudes-of-any-woman/}
}

@techreport{ArizeAI2024GenAISEC,
  title        = {The Rise of Generative AI in SEC Filings},
  institution  = {Arize AI},
  year         = {2024},
  month        = {July},
  url          = {https://arize.com/wp-content/uploads/2024/07/The-Rise-of-Generative-AI-In-SEC-Filings-Arize-AI-Report-2024.pdf},
  note         = {Accessed: 2025-05-20}
}

@techreport{BIS2024ExportControls,
  title        = {Commerce Strengthens Export Controls to Restrict China's Capability to Produce Advanced Semiconductors for Military Applications},
  institution  = {U.S. Department of Commerce, Bureau of Industry and Security},
  year         = {2024},
  month        = {April},
  note         = {Accessed: 2025-05-20}
}

@techreport{McAfee2023ArtificialImpostor,
  title        = {Beware the Artificial Impostor: A McAfee Cybersecurity Artificial Intelligence Report},
  institution  = {McAfee Corp.},
  year         = {2023},
  month        = {May},
  url          = {https://www.mcafee.com/content/dam/consumer/en-us/resources/cybersecurity/artificial-intelligence/rp-beware-the-artificial-impostor-report.pdf}
}

@online{EuropeanParliament2023EUAIAct,
  author  = {{European Parliament}},
  title   = {{EU AI Act: first regulation on artificial intelligence}},
  year    = {2023},
  url     = {https://www.europarl.europa.eu/topics/en/article/20230601STO93804/eu-ai-act-first-regulation-on-artificial-intelligence},
  note    = {Accessed: 2025-05-19}
}

@misc{fbi2023sextortion,
  title={International Law Enforcement Agencies Issue Joint Warning about Global Financial Sextortion Crisis},
  author={{Federal Bureau of Investigation}},
  year={2023},
  month={feb},
  note={Accessed: 2025-05-16}
}

@misc{lohn2025impactaicyberoffensedefense,
      title={The Impact of AI on the Cyber Offense-Defense Balance and the Character of Cyber Conflict}, 
      author={Andrew J. Lohn},
      year={2025},
      eprint={2504.13371},
      archivePrefix={arXiv},
      primaryClass={cs.CR},
      url={https://arxiv.org/abs/2504.13371}, 
}

@misc{agarwal2023llm,
  title={{LLM} Inference Performance Engineering: Best Practices},
  author={Agarwal, Megha and Qureshi, Asfandyar and Sardana, Nikhil and Li, Linden and Quevedo, Julian and Khudia, Daya},
  organization={Databricks},
  year={2023},
  month={oct},
  howpublished={\url{https://www.databricks.com/blog/2023/10/12/llm-inference-performance-engineering-best-practices.html}},
  note={Accessed: 2025-05-16}
}

@misc{aspen2024cyber,
  title={Envisioning Cyber Futures with {AI}},
  author={{Aspen U.S. Cybersecurity Group}},
  organization={Aspen Digital, Aspen Institute},
  year={2024},
  month={jan}
}

@article{bastos2019brexit,
  title={The Brexit Botnet and User-Generated Hyperpartisan News},
  author={Bastos, Marco T. and Mercea, Dan},
  journal={Social Science Computer Review},
  volume={37},
  number={1},
  pages={38--54},
  year={2019},
  doi={10.1177/0894439317734156}
}

@article{bernardi2024societal,
  title={Societal Adaptation to Advanced {AI}},
  author={Bernardi, Jules and Mukobi, George and Greaves, Hilary and Heim, Lennart and Anderljung, Markus},
  journal={arXiv},
  year={2024},
  month={may},
  archivePrefix={arXiv},
  eprint={2405.10295},
  primaryClass={cs.AI},
  doi={10.48550/arXiv.2405.10295}
}

@article{bruno2022brexit,
  title={Brexit and Bots: Characterizing the Behaviour of Automated Accounts on Twitter During the {UK} Election},
  author={Bruno, Matteo and Lambiotte, Renaud and Saracco, Fabio},
  journal={EPJ Data Science},
  volume={11},
  number={1},
  pages={17},
  year={2022},
  doi={10.1140/epjds/s13688-022-00346-2}
}

@article{naumov2019deep,
  title={Deep Learning Recommendation Model for Personalization and Recommendation Systems},
  author={Naumov, Maxim and others},
  journal={arXiv},
  year={2019},
  archivePrefix={arXiv},
  eprint={1906.00091},
  primaryClass={cs.IR},
  doi={10.48550/arXiv.1906.00091}
}

@techreport{Netwrix2024HybridSecurity,
  title        = {2025 Hybrid Security Trends Report},
  author       = {{Netwrix}},
  year         = {2024},
  institution  = {Netwrix Corporation},
  address      = {Irvine, CA},
  type         = {Technical Report},
  month        = {April},
  url          = {https://www.netwrix.com/2025-hybrid-security-trends-report.html},
  note         = {Accessed: May 21, 2025},
}

@article{roman2024proactive,
  title={Proactive Detection of Voice Cloning with Localized Watermarking},
  author={Roman, Robin San and Fernandez, Pierre and D\'efossez, Alexandre and Furon, Teddy and Tran, Tuan and Elsahar, Hady},
  journal={arXiv},
  year={2024},
  archivePrefix={arXiv},
  eprint={2401.12345},
  primaryClass={cs.SD},
  doi={10.48550/arXiv.2401.12345}
}

@article{sastry2024computing,
  title={Computing Power and the Governance of Artificial Intelligence},
  author={Sastry, Girish and Heim, Lennart and Belfield, Haydn and Anderljung, Markus and Brundage, Miles and Hazell, Julian and O'Keefe, Cullen and others},
  journal={arXiv},
  year={2024},
  archivePrefix={arXiv},
  eprint={2402.12345},
  primaryClass={cs.AI},
  doi={10.48550/arXiv.2402.12345}
}

@misc{slashnext2023phishing,
  title={The State of Phishing Report},
  author={{SlashNext}},
  year={2023},
  month={oct}
}

@misc{ubiquiti2015form8k,
  title={Form 8-K},
  author={{Ubiquiti Networks, Inc.}},
  year={2015},
  month={aug},
  day={4}
}

@incollection{unver2023strategic,
  title={The Strategic Logic of Digital Disinformation: Offence, Defence and Deterrence in Information Warfare},
  author={Unver, H. Akin and Arhan, S. Ertan},
  booktitle={Routledge Handbook of Disinformation and National Security},
  pages={192--207},
  publisher={Routledge},
  year={2023}
}

@misc{wayback2024openai,
  title={{OpenAI} Pricing},
  author={{Wayback Machine}},
  year={2024},
  howpublished={\url{https://web.archive.org/web/20230416151950/https://openai.com/pricing}},
  note={Accessed: 2024-08-01}
}

@article{zhou2023dont,
  title={Don't Make Your {LLM} an Evaluation Benchmark Cheater},
  author={Zhou, Kun and Zhu, Yutao and Chen, Zhipeng and Chen, Wentong and Zhao, Wayne Xin and Chen, Xu and Lin, Yankai and Wen, Ji-Rong and Han, Jiawei},
  journal={arXiv},
  year={2023},
  archivePrefix={arXiv},
  eprint={2301.12345},
  primaryClass={cs.CL},
  doi={10.48550/arXiv.2301.12345}
}
\bibliographystyle{plainnat}

\newpage
\appendix
\onecolumn
\section{HuggingFace Leaderboard data extraction}
\label{appendix_a}

The HuggingFace Open LLM leaderboard repository was cloned to access the full set of benchmark results. This repository contains information on each model, such as benchmark scores and model size; however, it does not contain all of the information needed to produce Fig \ref{fig1}. 

For example. the leaderboard repository does contain information on when the evaluation was performed, but it does not specify when the evaluated model was created, a date relevant to the discussion in Section 1. Given the leaderboard was created in mid-2023, there is a lag between evaluation date and creation date for models developed prior to this. For example, a model trained and released in 2022 could have been evaluated in late 2023.

To extract model creation date, we programmatically visit each model repository page. We scan through the commit history and identify the first commit where a ‘.config.json’ file was uploaded, a trigger event we assume indicates a training run has occurred. Of course, it is possible that a ‘config.json’ file was uploaded in the absence of a training run, leading to errors in our extracted creation dates. To verify the results, we manually compare our extracted creation dates to public model release announcements for the popular Llama, Mistral, and Phi-2 models. In all cases, we find our extracted date is consistent with the public announcement date, indicating our method, at least on this subset, performs well. 

Of course, if there is a significant gap between when a model was initially created and when it was evaluated, it is possible the model may have undergone many interim revisions – meaning the capabilities the model was evaluated at may not be the capabilities it possessed upon initial release. Given we are interested in evaluating open-source capabilities available at a given point in time, these scenarios could introduce errors into our findings. For example, suppose a model was created in 09/2022 but wasn’t added to the HuggingFace Leaderboard until 09/2023. Assume the model only offered an  $\alpha$=0.5 but was revised in 06/2023 to obtain an  $\alpha$=0.75 and re-uploaded to the same repository. In this scenario, our methodology would incorrectly associate a LLM capability of $\alpha$=0.75 with a date of 09/2022. 

We do not perform any additional evaluations to assess the severity of this reporting error. However, we speculate such errors are likely minimal. Substantial model revisions are time-consuming and expensive. If an AI lab performed such revisions, they would likely release them as a separate model rather than merely updating a model within an existing repo (i.e. Llama3 was created instead of Llama2 being updated), a scenario that would not produce an error within our methodology. 

\section{Compute profiling}
\label{appendix_b}

\begin{figure}
\centering
\includegraphics[width=0.5\linewidth]{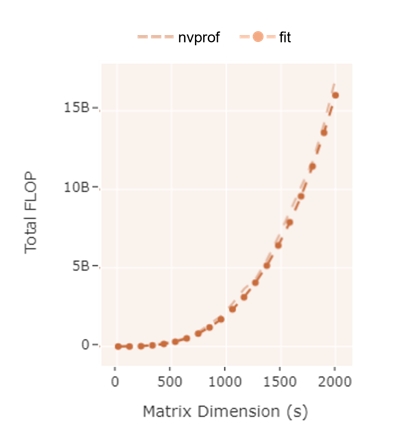}
\caption{The number of FLOPs required to execute a matrix multiplication are measured by a nvprof profiler. We then compare these measurements to well-established theoretical estimates, demonstrating good agreement and suggesting our profiler is working correctly.}
\label{appendix_b1}
\end{figure}

Using the nvprof tool, we profiled the GPU memory and compute requirements of a set of audio, video, and text models deployed on a NVIDIA V100 accelerator. Each profile was extracted by writing scripts to import relevant models from the transformers library, deploy them in inference, and record measurements on inference commands. We evaluated the accuracy of our profiler by using nvprof to measure the total FLOP of a single square matrix multiplication operation across a set of matrices of different dimensions and comparing the results to those from a well-established theoretical formula, observing good agreement (Fig. \ref{appendix_b1}). 

For image generation, we focused on the StabilityAI diffusion model (base and refiner versions); for audio modeling, we focused on Meta’s mms-tts-eng text-to-speech model and OpenAI’s Whisper speech-to-text model; and for text generation, we focused on Microsoft’s Phi-2 model and MistralAI’s Mistral-7B model. These models were selected by jointly considering their HuggingFace ‘trending’ status, number of downloads, and ‘like’ metrics – signs of how popular the models are within the developer community. 

We explored how both the memory and compute requirements for each model changed under different input/output conditions. For the text generation models, we measured how FLOP requirements changed as the inference batch size and number of output token generations were increased, assuming an input context length of 150 tokens. For image models, we investigated how both FLOP and memory requirements changed as image dimension, batch size, and number of de-noising steps were increased. Lastly, for audio models we measured how compute requirements changed as both the length of transcribed audio (speech-to-text) and number of converted tokens (text-to-speech) increased. The results are summarized in Figure \ref{appendix_b2} below.

\begin{figure}
\centering
\includegraphics[width=1.0\linewidth]{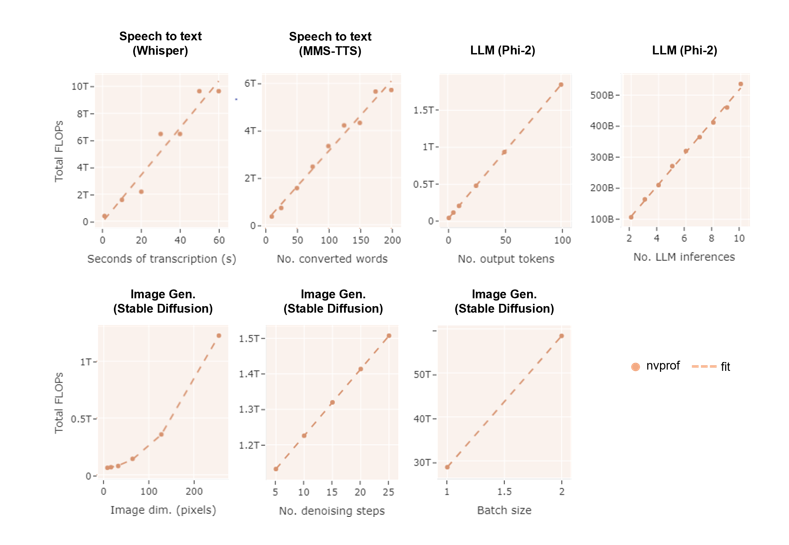}
\caption{Calibration curves that measure the amount of FLOPs required to run audio, image, and text generation models in inference under a variety of settings including batch size, image dimension, output token length, and others. All measurements were performed through the nvprof profiling tool embedded within a V100 GPU.}
\label{appendix_b2}
\end{figure}

\section{Description of case studies}
\label{appendix_c}

\textit{Ubiquiti Networking Spear-phishing (2015, $\sim$\$50M in losses)}
In 2015, Ubiquiti was victim to an employee spear-phishing campaign that inflicted over \$46M in company losses \citep{ubiquiti2015form8k}. The attackers impersonated employees while making fraudulent wire transfer requests to other unsuspecting employees. This event falls within a more general category of business email compromise (BEC) fraud that was responsible for \$2.9B in losses in 2023 alone and is becoming increasingly more common with the advent of generative AI \citep{slashnext2023phishing}.

In our simulation, we estimated the compute required for a LLM to generate synthetic emails to $\sim$1000 employees, a workforce consistent to that of Ubiquiti in 2015. We assume each employee is sent approximately 10 emails over a two-week period, a timescale also similar to that reported in the Ubiquiti campaign. Within this exercise, we assume an attacker already has access to the corporate email accounts needed for the BEC scheme, and a LLM is solely responsible for orchestrating communications with the targeted employees.

\textit{DeepNude (2019, 100,000+ model downloads)}
Released on GitHub in 2019, DeepNude is a generative model that ingests real images of clothed individuals and produces synthetic nude images of these individuals. Attackers have leveraged this model within sex-tortion campaigns where they produce synthetic nude images of victims and threaten to release such images publicly unless the victim transfers money or other assets to the attacker. Within the first week of release, the model was downloaded over 95,000 times and several mirror downloads sites came online after the initial code repository was shut down. Since then, the number of sextortion campaigns has skyrocketed, with the FBI declaring a “global financial sextortion crisis” in 2023.

In our simulation of this scenario, we estimated the compute resources needed to generate $\sim$100,000 synthetic images over a roughly one-month time period via open-source generative diffusion models.

\textit{Voice-cloning fraud (2023, affecting $\sim$25\% of Americans)}
Voice cloning has emerged as a concerning vehicle for financial fraud. In a recent report, McAfee (2023) reported that 25\% of surveyed U.S. adults have either experienced or known someone who has experienced an AI voice scam. While many different fraud schemes exist, in one version, attackers first collect publicly available audio samples from a victim through social media profiles or other online sources. They then fine-tune an audio generation model to recite a specified piece of text in a voice nearly indistinguishable from their victim. Leveraging the cloned voice, attackers will send distressed voicemails to their victim’s close social contacts, requesting money transfers to the attacker’s account. Attackers may also choose to have an AI model engage in real-time conversation rather than merely send voicemails. Similarly, attackers may not need to fine-tune an audio model on samples from their victim's close social contacts. 

We simulated the computational profile for a fully automated form of voice cloning in which a LLM generates real time conversation responses which are converted into synthetic audio samples by an audio generation model in real time. Such a system would need to convert a victim’s audio into text, process this text into a response, and convert this response into a cloned audio sample within a seconds-long timescale. We consider each of these steps when calculating the computational profile presented in Section \ref{threat} but we do not profile any steps related to fine-tuning of the audio model itself since this step is not required in all fraud schemes. 

\textit{Brexit botnet (2017, ~65,000 Tweet disinformation campaign)}
Online botnets have been leveraged within influence operations to sow social discord, amplify disinformation, and engender radicalization \citep{bruno2022brexit}. In our simulation exercise, we profiled the computational load required for a LLM to execute an X (formerly Twitter) disinformation operation similar in scale to the well-studied 2017 Brexit botnet campaign \citep{bastos2019brexit}. Within this campaign, 13,000 bot accounts posted roughly $\sim$65,000 divisive tweets over a two-week period to steer public opinion surrounding the Brexit referendum vote.

\section{Compute profiling of case studies}
\label{appendix_d}

We used the following formula to estimate the compute profile for each synthetic media campaign:

\[
C_k = \frac{1}{T} \left( \sum_{j=1}^{N_I} p_{I,k}(d_j, m_j) + \sum_{j=1}^{N_{TTS}} p_{TTS,k}(t_{j,TTS}) + \sum_{j=1}^{N_{STT}} p_{STT,k}(s_j) + \sum_{j=1}^{N_L} p_{L,k}(t_{j,L}, n_j) \right)
\]
where \( k \in \{F, M\} \), \( C_F \) is estimated processing power (FP32 FLOPS); \( C_M \) is memory bandwidth (GB/s); \( N_i \) refers to the number of generation steps for each model with \( i \in \{I, TTS, STT, L\} \) corresponding to image, text-to-speech, speech-to-text, and language models, respectively; \( p_{I,k}(d_j, m_j) \) is the processing power or memory transactions associated with producing an image of dimension \( d_j \) with \( m_j \) denoising steps; \( p_{TTS,k}(t_{j,TTS}) \) is the processing power or memory transactions associated with converting \( t_{j,TTS} \) tokens of text into speech; \( p_{STT,k}(s_j) \) is the processing power or memory transactions associated with converting \( s_j \) seconds of audio into text; and \( p_{L,k}(t_{j,L}, n_j) \) is the processing power or memory transactions associated with generating \( t_{j,L} \) tokens of text with an LLM of \( n_j \) parameters.

The functional forms of \( p_{I,k} \) were extracted either from the fits in Figure \ref{appendix_b2} of Appendix \ref{appendix_b} or theoretical estimates, when available. For each campaign, we calculated a distribution of \( C_k \) by sampling over a uniform range of \( T, N_I, N_{TTS}, N_{STT}, N_L, d_j, n_j, \) and \( t_j \). The ranges for these variables, displayed in Tabel \ref{tab:monte-carlo-sampling}, represent uncertainties in both the timeline and scale of each considered case study (number of social posts in a disinformation campaign, number of tokens in a spear-phishing email, etc.), as well as what model inference settings may be needed to replicate them (size of LLM, number of denoising steps within a StableDiffusion model, etc.). To produce the most conservative uncertainty bounds, for each Monte-Carlo trial, we fixed the value of all sampled variables as opposed to resampling the values for each summation term (i.e., within a single Monte-Carlo trial, either a 2B LLM or a 7B LLM was assumed but trials where a mixture of LLMs were used were not considered).

\begin{figure}
\centering
\includegraphics[width=0.8\linewidth]{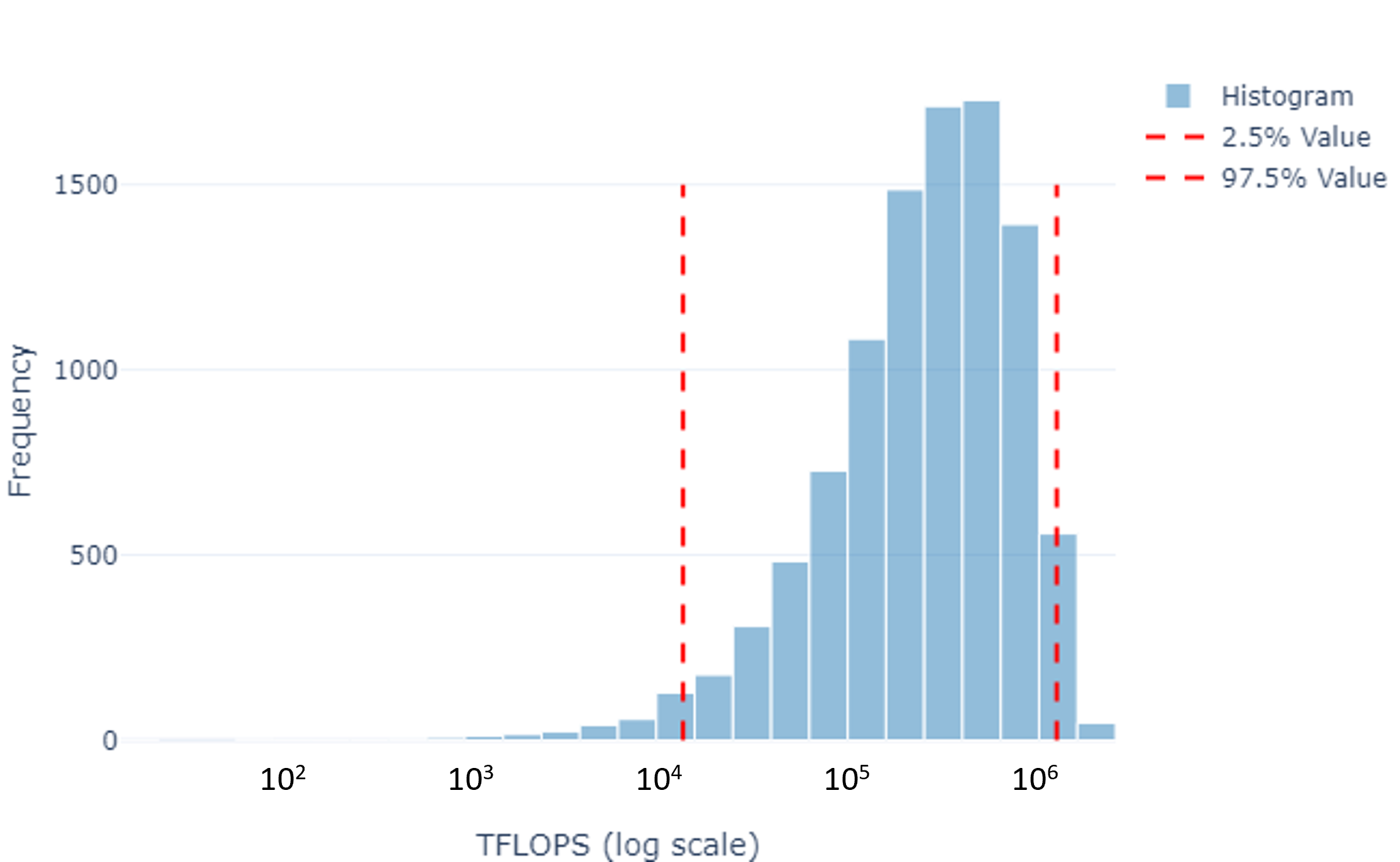}
\caption{Compute budget for the botnet scenario across a set of Monte-carlo simulations. Dashed lines represent the 2.5\% and 97.5\% percentile values of the distribution.}
\label{botnet-sim}
\end{figure}

In Figure \ref{botnet-sim}, we display the distribution of simulated compute across all trial runs for the botnet disinformation campaign as an illustrative example. For each social harm campaign, we report the 2.5\% and 97.5\% percentile values of the simulated distributions as lower and upper bounds for their associated performance metrics, respectively. 

There is also additional uncertainty related to unknown computational tasks that may be required to execute each campaign that are not considered by our methodology. For example, before generating a spearphishing email, an LLM might first scrape the internet for personal information on a victim employee. This background research could require an LLM to execute Google search queries, extract relevant content from social media profiles, and summarize retrieved information for subsequent queries. Similarly, each attack may be leveraged with a reasoning models, such as DeepSeek r1 or o1, which generate additional reasoning tokens along with their final output \citep{guo2025deepseek}. To account for unconsidered computational costs such as these, we conservatively add an additional order of magnitude to the memory bandwidth and processing power upper bounds for all considered campaigns.

On the other hand, plenty of research is exploring how AI models can be run in more compute-efficient ways. For example, FlashAttention leverages efficient caching of variables to reduce memory bandwidth \citep{dao2022flashattention}, while mixture-of-expert modeling drives costs down by requiring fewer memory transactions per inference. To account for the existence of memory and compute optimizations not contained in our simulations, we conservatively also reduce the lower bounds for the estimated compute profile by a factor of ten.

\begin{table}[t]
\caption{Monte-Carlo sampling parameters for compute profiles}
\label{tab:monte-carlo-sampling}
\centering

\small 
\setlength{\tabcolsep}{4pt} 
\renewcommand{\arraystretch}{1.15} 

\begin{tabular}{%
>{\raggedright\arraybackslash}p{0.15\textwidth} 
>{\raggedright\arraybackslash}p{0.23\textwidth} 
>{\raggedright\arraybackslash}p{0.09\textwidth} 
>{\raggedright\arraybackslash}p{0.20\textwidth} 
>{\raggedright\arraybackslash}p{0.29\textwidth} 
}
\toprule
\textbf{Campaign} &
\textbf{Profiled models} &
\textbf{$T$} &
\textbf{No. synthetic generations} &
\textbf{Generation parameters} \\
\midrule

Spear-phishing & 
Microsoft Phi-2B, Mistral 7B & 
15--30 days & 
NI: 2 & 
$d_j$: 512--1024 \\

& & & 
NTTS: 0 & 
$m_j$: 5--50 \\

& & & 
NSST: 0 & 
$t_{j,\mathrm{TTS}}$: --- \\

& & & 
NL: 2,500--25,000 & 
$s_j$: --- \\

& & & &
$n_{j,\mathrm{TTS}}$: 2--30 B params. \\

& & & &
$t_{j,L}$: 100--500 tokens \\[0.35em]

Brexit botnet & 
Microsoft Phi-2B, Mistral 7B & 
15--30 days & 
NI: 0 & 
$d_j$: --- \\

& & & 
NTTS: 0 & 
$m_j$: --- \\

& & & 
NSST: 0 & 
$t_{j,\mathrm{TTS}}$: --- \\

& & & 
NL: 50,000--100,000 & 
$s_j$: --- \\

& & & &
$n_{j,\mathrm{TTS}}$: 2--30 B params. \\

& & & &
$t_{j,L}$: 50--100 tokens \\[0.35em]

Voice cloning & 
Microsoft Phi-2B, Mistral 7B, Facebook MMS-TTS, \mbox{OpenAI Whisper} & 
1--2 s & 
NI: 0 & 
$d_j$: --- \\

& & & 
NTTS: 1 & 
$m_j$: --- \\

& & & 
NSST: 1 & 
$t_{j,\mathrm{TTS}}$: 3--5 tokens \\

& & & 
NL: 1 & 
$s_j$: 2--10 s \\

& & & &
$n_{j,\mathrm{TTS}}$: 2--30 B params. \\

& & & &
$t_{j,L}$: 10--30 tokens \\[0.35em]

DeepNude & 
Stable Diffusion XL + \mbox{Stable Diffusion Refiner} & 
50--100 days & 
NI: 0 & 
$d_j$: 512--1024 \\

& & & 
NTTS: 0 & 
$m_j$: 5--50 \\

& & & 
NSST: 0 & 
$t_{j,\mathrm{TTS}}$: --- \\

& & & 
NL: 75,000--150,000 & 
$s_j$: --- \\

& & & &
$n_{j,\mathrm{TTS}}$: --- \\

& & & &
$t_{j,L}$: --- \\
\bottomrule
\end{tabular}
\end{table}

\newpage
\section{Simulation of non-nefarious workloads}
\label{appendix_e}

We profiled the total FLOP required for each workload, either by inferring profiles from hardware reports or directly simulating these workloads within a test environment, similar to Section \ref{threat}. When using hardware reports, we assume the compute required for a workload is bounded by the 10 – 90 \% performance limits of the GPU on which it was executed \citep{agarwal2023llm}. As AI models are generally designed to run efficiently on the hardware that supports them, our underlying assumption is the computational requirements of a given AI job often hovers close to, or at least within an order of magnitude of, the maximum workload the GPU can execute.
The autonomous vehicle recognition profile was calculated assuming object detection is performed with a YOLOv8m model (80B FLOP/inference) over eight camera sensors operating at a frame-per-second. Here, we do not consider any computational loads due to trajectory planning, 3D reconstruction, or other tasks, meaning our estimates are an extremely conservative lower bound. Additional details on the profiling for each scenario are provided below:

\begin{itemize}
\item The ALPHAFOLD profile was calculated by assuming 10-90\% maximum operating efficiency of their reported cluster of 40 NVIDIA RTX6000 GPU inference GPUs. 
\item The call center transcription was profiled assuming the Whisper compute requirements outlined in Appendix \ref{appendix_d} with an inference load of (8 hours) x (3600 seconds/hour) x (20 employees) = 572,000 seconds of total transcribed audio. 
\item The spam detection profile also followed a methodology similar to Appendix  \ref{appendix_d}, with the LLM size fixed at $\sim$500M parameters (chosen to match a NVIDIA BERT-based spam detection system) and an inference load of (1,000 employees) x (100 emails/employee) x (100 tokens/email) = 800,000 processed tokens. 
\item The deep learning recommendation system was profiled by replicating the Criteo Advertising Challenge benchmark training script published by an open source research group \citep{naumov2019deep}. We leveraged nvprof to measure the FLOP of a single training mini-batch and up-scaled this value up to simulate the total required FLOP for a full 300-epoch, 1000 mini-batch/batch training run. 
\end{itemize}

In Table \ref{tab:compute-profile}, we display the extracted compute profile of each considered workload. 

\begin{table}[t]
\caption{Compute profiles of non-nefarious AI use cases}
\label{tab:compute-profile}
\centering
\setlength{\tabcolsep}{4pt} 
\small 
\renewcommand{\arraystretch}{1.3} 
\begin{tabular}{p{0.2\textwidth}p{0.25\textwidth}p{0.25\textwidth}p{0.15\textwidth}}
\hline
Compute workload & Model description & Workload description & TFLOP estimate \\[2pt]
\hline
Autonomous-vehicle object recognition & YOLOv8a object detector for autonomous navigation & 4\,h of vehicle navigation & $10^4 - 10^5$ \\[6pt]
Scientific computing & AlphaFold protein-structure prediction & Structure prediction for 559 proteins & $10^5 - 10^6$ \\[6pt]
Call-center transcription & Whisper speech-to-text for a 20-seat call center & 8\,h of call-center operation & $10^1 - 10^2$ \\[6pt]
Enterprise spear-phishing detection & Morpheus binary classifier for spear-phishing e-mails & 8\,h of e-mail screening in a 1000-person organization & $10^2 - 10^3$ \\[6pt]
Recommendation-engine training & Meta DLRM on the Criteo display-advertising benchmark & Single training run & $10^6 - 10^7$ \\[2pt]
\hline
\end{tabular}
\end{table}

\end{document}